\documentclass[fleqn,twoside]{article}
\usepackage{espcrc2}
\usepackage[dvips]{graphicx}
\usepackage{latexsym}
\usepackage{euscript}
\newcommand{\ifig}[1]{\includegraphics[height=8cm,width=8cm]{#1}}

\newcommand{\bc}{\begin{center}}
\newcommand{\ec}{\end{center}}
\newcommand{\be}{\begin{equation}}
\newcommand{\ee}{\end{equation}}
\newcommand{\bq}{\begin{quote}}
\newcommand{\eq}{\end{quote}}

\newcommand{\AmS}{{\protect\the\textfont2
  A\kern-.1667em\lower.5ex\hbox{M}\kern-.125emS}}
\title{SU(2) Glueballs, diquarks and mesons
in  dense matter}
\author{M.-P. Lombardo\address{INFN, Laboratori Nazionali di Frascati, 
\\via E. Fermi 40,
I00044 Frascati, Italy}, M. L. Paciello\address[ROMA1]{INFN, Sezione di 
Roma 1, \\
P.le A. Moro 2, I-00185 Roma, Italy}, 
S. Petrarca\address{Dipartimento di Fisica, Universit\`a di Roma ``La
Sapienza'', \\P.le A. Moro 2, I-00185 Roma, Italy}\addressmark[ROMA1], 
B. Taglienti\addressmark[ROMA1]}
\begin{document}
\begin{abstract}
We present preliminary results from a high statistics study of 2-color
QCD at low temperature and non-zero baryon density.
The simulations are carried out on a $6^3 \times 12$ lattice and use a 
standard
hybrid molecular dynamics algorithm for staggered fermions for two values
of quark mass.
Observables include glueball correlators evaluated via a multi-step
smearing procedure as well as scalar and vector mesons and diquarks.
\vspace{1pc}
\end{abstract}
\maketitle

\section{Motivation}

The SU(2) QCD lattice theory has been recently reconsidered as a good 
starting point
in order to get some insight into QCD phenomena at chemical 
potential $\mu$ and
temperature different from zero. The main difficulty of the complex 
action    once     $\mu \neq 0$
is avoided adopting  the color SU(2) group as in this 
case the determinant is real and 
standard hybrid Monte Carlo simulations
are possible. As a general reference see \cite{Sinc}.

In the following we discuss the scalar gluonic correlators  constructed 
by the smearing technique \cite{smearing}
in order to excite glueballs and compare their behavior with
that of scalar mesons as a function of baryon density.
The study of the pattern of chiral symmetry in the vector fermionic 
sector and
of   vector condensation signals is also addressed.

\section{The simulation}

The simulations were performed on $6^3 \times 12$ lattice using a standard
hybrid molecular dynamics algorithm for staggered fermions \cite{hklm} at 
$\beta=1.3$  for two
quark masses: $m=0.05$ and $m=0.07$ and four values of the chemical 
potential: $\mu= 0.0, 0.2, 0.4, 0.6$.
 We performed $400000$ steps saving configurations
every $10$ steps. We also performed $30000$ steps for $m=0.07$ and 
$\mu=0.25, 0.30, 0.32, 0.35, 0.90$.

The glueball correlation functions have been
obtained using the smearing procedure \cite{smearing}
trying six values of the smearing weight: $w= 0.025, 0.05, 0.07, 0.1, 
0.2, 0.3$ repeated
recursively $1,2,3,4$ times (smearing steps). Our analysis is in 
progress  and the data shown in the following are to be considered
preliminary, more details will be given in \cite{noi}. See ref. 
\cite{ishii} for
recent result on glueballs at finite temperature.
In Figs. 1
and ~\ref{fig:gmass} we show the good
fluctuation reduction obtained by the smearing procedure;
the glueball correlators and masses are plotted at $\mu \ne 0$ as a 
function of
a combined smearing variable (steps times weight).
The value of masses are obtained as logarithm of the glueball 
correlation ratios
at different distances.\\
\begin{figure}[h]
\bc
\ifig{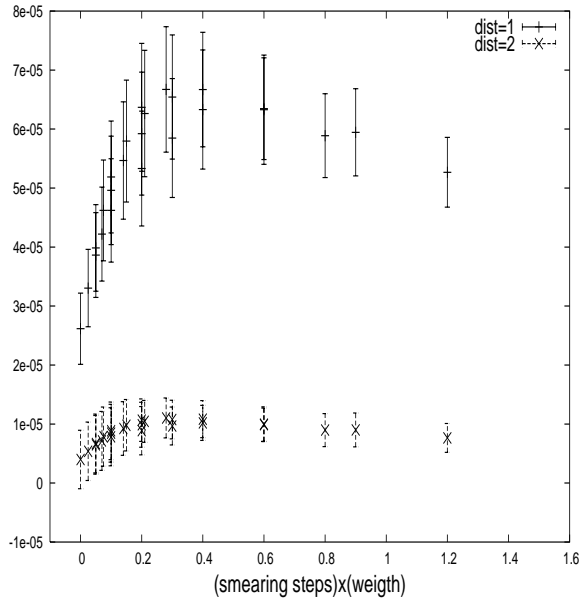}
\caption{ Glueball correlations at distance 1 and 2
as a function of the product smearing (steps $\cdot$  weight) for $\mu=0.2$
and $m=0.07$.
The smearing strongly reduces the fluctuations.}
\ec
\label{fig:smea}
\end{figure}
\begin{figure}[h]
\bc
\ifig{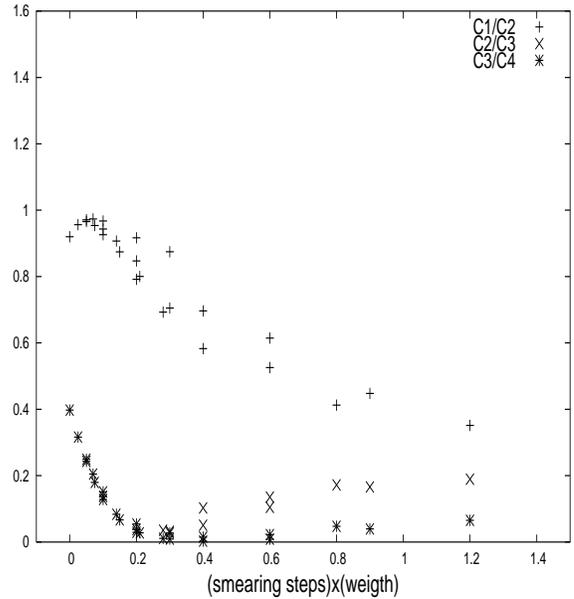}
\caption{Glueball masses
as a function of the product smearing (steps $\cdot$  weight) for $\mu=0.6$
and $m=0.05$.
}
\ec
\label{fig:gmass}
\end{figure}
\begin{figure}[h]
\bc
\ifig{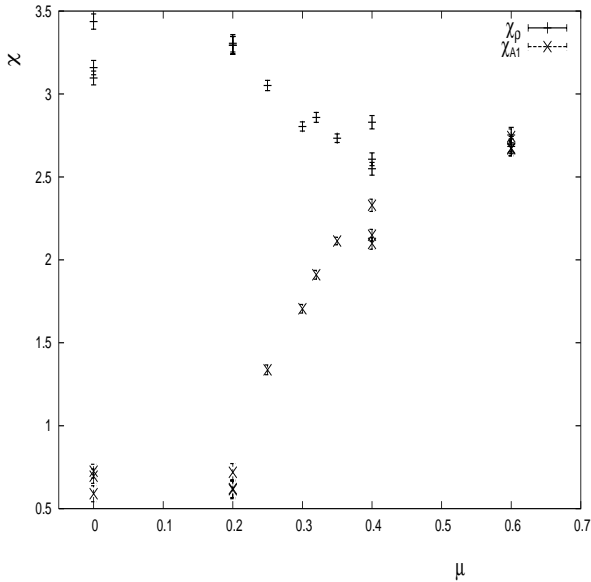}
\caption{Susceptibility for $\rho$ and $A1$ as
function of the chemical potential $\mu$, $m = 0.07$.}
\ec
\label{sushi}
\end{figure}
\begin{figure}[h]
\bc
\ifig{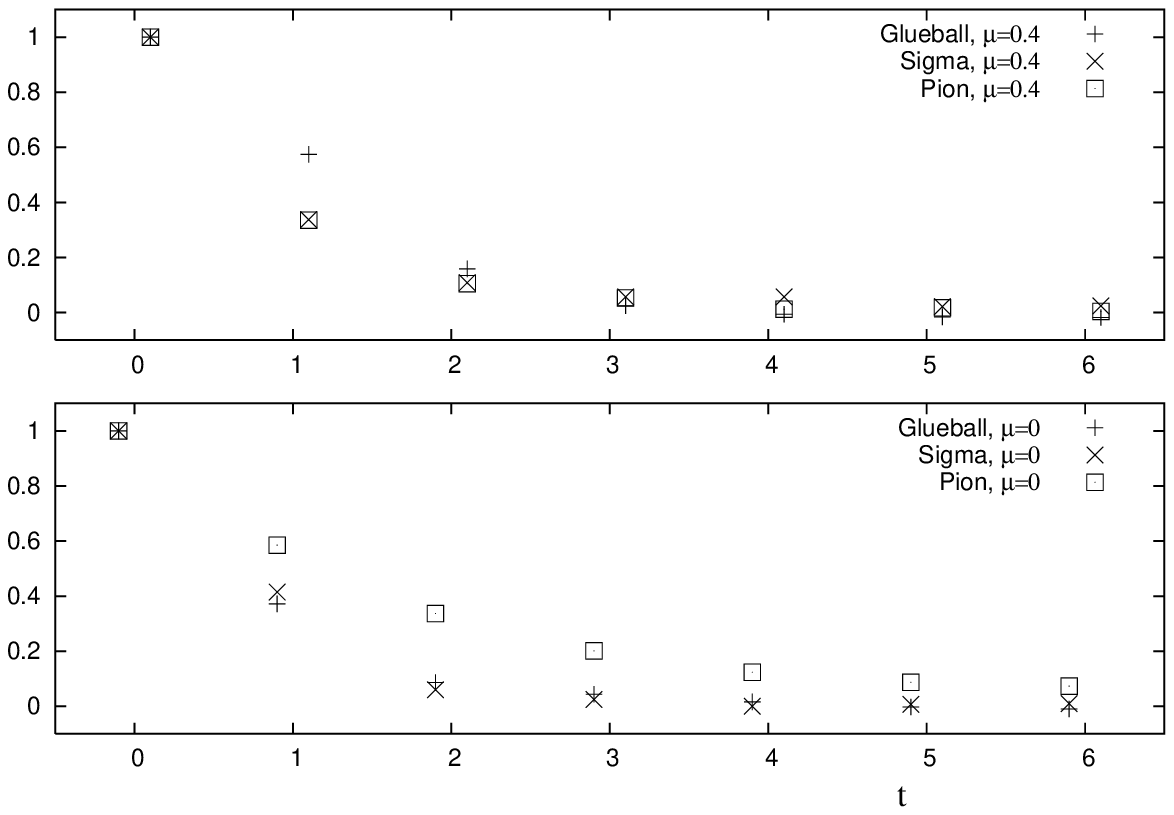}
\caption{Level ordering at $\mu=0$ and $\mu=0.4 $, $m_q = 0.05$:
$G(t)/G(0)$ is plotted against the Euclidean time $t$.
At $\mu=0$: $m_\pi < {m_\sigma \sim m_{G^{00^{+}}}}$;
at $\mu=0.4$: $ m_{G^{00^{+}}}< {m_\sigma \sim m_\pi}$.}
\ec
\label{scalar}
\end{figure}
The general features of the results are monitored by the behavior of
the particle number matrix element $< {\bar{ \psi} \gamma_0 \psi}>$ and
of the chiral condensate $<{{\bar{ \psi}} \psi}>$ as a function
of the chemical potential $\mu$.
We find a clear signal of  phase transitions  at $\mu_c = 0.25 $ for 
$m=0.05$,
and $\mu_c = 0.3 $ for $m=0.07$ in agreement with \cite{hklm}.

In Fig. 3 we show the behavior of the chiral susceptibility for the 
$\rho$
and the $A_1$ as a function of $\mu$. After the phase transition the two
states become degenerate, consistently with a vanishing chiral
condensate.  However, the behavior of our staggered propagators at 
finite density is different from the free-like one
observed at high temperature,
confirming the observations of \cite{gise}.

The trend of  Fig. 3 suggests
that the $\rho$ becomes heavier, while the $A1$ becomes lighter
with density.  The propagator themselves do not show any dramatic 
change, but, again their behavior suggests an 
heavier $\rho$ in the dense phase. Note that a recent study with
Wilson fermions \cite{Muroya:2002ry} reports instead
a ligher $\rho$ in the medium. As spectrum's qualitative features
might well depend on the mass range,  and on lattice artifacts, 
more work is needed  before
drawing general conclusions on the behavior of vector mesons
in dense matter.

The vector diquark correlators 
do not show any hint of the predicted vector
condensation \cite{San}, but the saturation effects might well
hide this phenomenon (
the $\rho$ mass at zero chemical potential was estimated to be
$m_\rho \simeq 1.8$, hence the threshold $\mu_v = m_{\rho}/2$ is 
well inside the saturation region).

A sample of the results for the scalar sector is given in Fig.\ 4:
the lower diagram shows the propagators, normalised to their
zero distance value, for the pion, sigma, and scalar glueball
in the normal phase; the pion (the Goldstone meson) is
the lightest particle. In the superfluid phase (upper diagram)
the sigma meson and the pion are degenerate, and heavier than 
the glueball. In addition,
by contrasting lower and upper diagram we note that the 
glueball is lighter in the medium.

\section{Summary}
It is possible to measure glueball states at $\mu \neq 0$.
 The smearing procedure \cite{smearing}, 
already tested in the theory with $\mu = 0.0$, 
increases very much
the signal. The correlators at high density show  the usual
behavior.\\
In the vector sector we see the expected  $\rho$ - $A_1$ degeneracy 
at high density and  we find a
behavior clearly different from the high temperature one.\\
The $\rho$ appears to be heavier in dense matter, 
at least for $m_q = 0.07$.\\
We find a  different level ordering in the scalar sector
after the phase transition:
from $m_\pi < {m_\sigma \sim m_{G^{00^{+}}}}$ at  $\mu=0.0$ to
$ m_{G^{00^{+}}}< {m_\sigma \sim m_\pi}$ at $\mu=0.4$

\end{document}